\documentclass[aps,prl,twocolumn,floats,showpacs,amsmath,amssymb,floatfix]{revtex4}
\usepackage[dvips]{color}
\usepackage{graphicx}
\usepackage{epsfig}
\usepackage{dcolumn}
\usepackage{bm}
\usepackage{amsmath}
\usepackage{amssymb}

\begin{document}
\title{Thermoelectric Transport through a Quantum Dot:\\
Effects of Kondo Channels Asymmetry}
\date{\today}

\author{T. K. T. Nguyen, M. N. Kiselev, and  V. E. Kravtsov}

\affiliation{The Abdus Salam International Centre for Theoretical
Physics, Strada Costiera 11, I-34151, Trieste, Italy}

\begin{abstract}
We consider effects of magnetic field on the thermopower and
thermoconductance of a single-electron transistor based on a quantum
dot strongly coupled to one of the leads by a single-mode quantum
point contact. We show appearance of two new energy scales:
$T_{min}$$\sim$$|r|^2$$E_C$$($$B$$/$$B_C$$)$$^2$ depending on a
ratio of magnetic field $B$  and the field $B_C$ corresponding to a
full polarization of point contact and $T_{max}$$\sim$$|r|^2$$E_C$
depending on a reflection amplitude $r$ and charging energy $E_C$.
We predict that the behavior of thermoelectric coefficients is
consistent with the Fermi-liquid theory at temperatures
$T$$\ll$$T_{min}$, while crossover from Non-Fermi-liquid regime
associated with a two-channel Kondo effect to Fermi-liquid
single-channel Kondo behavior can be seen at
$T_{min}$$<$$T$$<$$T_{max}$.
\end{abstract}
\pacs{
  73.23.Hk,
  73.50.Lw,
  72.15.Qm,
  73.21.La
 }
\maketitle \vspace*{-5mm} The thermoelectric transport through
nanostructures is a subject of extensive experimental
\cite{Staring93,Dzurak97,PhysRevLett.95.176602,PhysRevB.75.041301(R)}
and theoretical
\cite{PhysRevB.46.9667,MA_theory,PhysRevB.65.115332,EuroPhysLett.56.576}
studies. One of particularly interesting questions is related to the
thermoelectric properties of Quantum Dots (QD) in a Coulomb Blockade
(CB) regime. While the experimental behavior of thermoelectric
transport through weakly coupled QD-devices is well described
theoretically \cite{PhysRevB.46.9667,MA_theory}, the regime of
strong coupling of the QD to the reservoirs is far from being
understood.

The strong enhancement of thermopower is important for
nano-technology applications \cite{PhysRevB.75.041301(R)}. It
provides a challenge for both experimental fabrication of devices
with high thermo-conductance and theoretical suggestions for
efficient mechanisms of a heat transfer. On one hand, the
nano-technologies offer fascinating tunability of single electron
transport, while most parameters can be changed continuously by
applying gate voltage, external electric and magnetic fields etc. On
the other hand, the nano-devices efficiently operate in strong
coupling regime where effects of strong electron correlations can be
viewed as a prominent mechanism for the thermoelectric coefficients
enhancement. In particular, the Kondo effect is known as a tool for
strong intensification of electric transport through single electron
transistor (SET). Moreover, by increasing the number of channels one
can fine tune SET to a Non-Fermi liquid (NFL) regime. The NFL
behavior is however illusive being very sensitive to variation of
external parameters since channel symmetry is generally unprotected
by conservation laws. Thus, by perturbing the SET with external
fields one creates a mechanism of restoration of Fermi-liquid (FL)
transport. These effects are known in the theory of electric
transport \cite{Goldhaber}. In this Letter we present a theory of an
interplay between NFL and FL strong coupling regimes in
thermoelectric transport through the nano-structures.

Typical experimental setup \cite{PhysRevLett.95.176602} for
measuring the thermopower $S$$=$$-$$\Delta V_{th}$$/$$\Delta T$ is
shown on Fig.1. The measurement of thermo-voltage $\Delta$$V_{th}$
provides independent information on the thermo-conductance $G_T$.
The temperature difference across the dot $\Delta$$T$ is controlled
by using a current heating technique. The differential conductance
$G$ is measured at variable gate voltages $V_g$. Similar to
differential conductance, the thermopower $S$$=$$G_T$$/$$G$ shows
the oscillations as a function of $V_g$. However, these oscillations
are not sinusoidal at the strong coupling regime. Moreover, no
relation analogous to Cutler-Mott formula \cite{Cutler69}
$S$$\sim$$\partial$$\ln G$$/$$\partial V_g$ exists in that limit
underlining importance of strong electron correlations.

Theory of thermopower (TP) of a CB - quantum dot has  been studied
in \cite{PhysRevB.46.9667} using a linear response theory in both
classical and quantum regimes and verified by experiment
\cite{Staring93, Dzurak97}.  For a week coupling of the dot to
reservoirs, there is no deviation from the theory ignoring effects
of interaction. By increasing the coupling of the SET to one (or
both) reservoirs one reaches the strong coupling regime where Kondo
physics becomes important for odd electron occupation number in the
dot. The thermoelectric transport through the SET in the Kondo
regime has been studied numerically in \cite{EuroPhysLett.56.576}.

Another approach to describe SET in the strong coupling regime has
been proposed in \cite{PhysRevB.51.1743,FM_theory,Flensberg}. The
theory of the TP of the SET designed as the quantum dot strongly
coupled to one of the leads through almost transparent Quantum Point
Contact (QPC) \cite{Houten92} and weakly coupled to second reservoir
has been constructed by Andreev and Matveev (AM) in
\cite{MA_theory}. In the model \cite{MA_theory} the dot was assumed
large enough to disregard the effects of finite mean-level spacing
$\delta$ \cite{cm1}. There is no odd-even effects related to the
occupation of the dot since the QD+QPC mimics a $\sigma$$=$$1$$/$$2$
quantum impurity. The transport of electrons is dominated by the
inelastic cotunneling at temperatures $T$$\ll$$E_C$. The two limits
were considered in \cite{MA_theory}: i) the electron spins are fully
polarized by strong external magnetic field $B$ \cite{com00} and ii)
the electrons spins are unpolarized, $B$$=$$0$. In the first limit,
corresponding to single channel Kondo (1CK) physics, the TP shows
sinusoidal oscillations as a function of the gate voltage $V_g$
having nodes both in Coulomb valleys ($N$ is integer) and peaks ($N$
is half-integer)
\begin{eqnarray}
S\sim \frac{1}{e}|r|\frac{T}{E_C}\sin(2 \pi N (V_g)). \label{sl1}
\end{eqnarray}
Here $N(V_g)$ is a dimensionless parameter which is proportional to
$V_g$.  The TP is a linear function of both reflection amplitude
$|r|$ and temperature. We refer to the linear-T dependence of TP as
the FL regime. However, in contrast to TP in bulk metals $S\sim
T/\epsilon_F$, the energy in denominator (\ref{sl1}) is much smaller
\cite{com} compared to Fermi energy $\epsilon_F$ (cf. \cite{Ong03})
thus enhancing the TP.
\begin{figure}[t]
\includegraphics[width=5.0cm,angle=0]{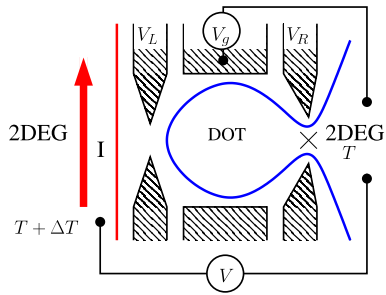}
\caption{(Color online) Experimental setup of SET in the strong
coupling regime (see text for the details). The arrow along left
lead stands for the electric current controlling the Joule heat.
Almost transparent QPC remaining at the reference temperature $T$ is
denoted by the cross.} \label{f.1}
\end{figure}

There are two important regimes for thermoelectric coefficients in
the limit of unpolarized electrons. First, at temperatures
$E_C|r|^2\ll T\ll E_C$ the TP oscillations are sinusoidal and
quadratic in the reflection amplitude $|r|$:
\begin{eqnarray}
S\sim \frac{1}{e} |r|^2\ln\left[\frac{E_C}{T}\right]\sin(2 \pi
N(V_g)). \label{sf1}
\end{eqnarray}
Second, at smaller temperatures $T\ll E_C|r|^2$, the TP oscillations
are non-sinusoidal
\begin{eqnarray}
S\sim \frac{1}{e}
|r|^2\ln\left[\frac{E_C}{T+\Gamma(V_g)}\right]\sin(2 \pi N(V_g))
f\left(\frac{\Gamma(V_g)}{T}\right), \label{sf2}
\end{eqnarray}
where $\Gamma(V_g) \sim E_C|r|^2\cos^2(\pi N(V_g))$ and $f(x)$ is
defined in \cite{com1}. The amplitude of $S$ scales as $S_{\max}\sim
e^{-1}|r|\sqrt{T/E_C}\ln(E_C/T)$. Thus, the TP is strongly enhanced
by the electron's correlations in the strong coupling regime. The
$\sqrt{T}\ln T$ scaling of $S$ - maximums is related to the nodes of
$\Gamma$ at the Coulomb peaks. The ``effective width" $\Gamma$ cuts
the log in (\ref{sf1}) everywhere away from the peaks. We refer to
$\sqrt{T}\ln T$ behavior of $S_{max}$ as the NFL regime. Such
scaling of the TP at $T$$\ll$$E_C$ is attributed to the two-channel
Kondo effect (2CK) \cite{NozBla80,Affleck91} where the channels
$\uparrow$ and $\downarrow$ scatter on $\sigma$$=$$1$$/$$2$ isospin
associated with left- and right- movers passing through the QD+QPC
\cite{com2}. However, as it is known \cite{NozBla80}, the 2CK strong
coupling fixed point is unstable. Therefore, any perturbation
resulting in the channel asymmetry becomes relevant
\cite{Affleck91}. In particular, effects of Zeeman smearing of
Coulomb staircase steps due to magnetic field driven asymmetry of
reflection amplitudes was studied in \cite{PhysRevB.64.161302} for
the almost open QD {\it capacitively} coupled to the gate.  Thus, as
it was shown in \cite{PhysRevB.64.161302}, magnetic field produces
relevant perturbation for AM model. Unstable nature of 2CK strong
coupling fixed point makes it very tricky to observe the
fingerprints of NFL in electron transport through nanostructures
\cite{Goldhaber}. We show in this Letter that the proximity to NFL
regime can be seen in the thermoelectric transport tuned by external
magnetic field and address the question how the thermoelectric
coefficients evolve in the presence of finite magnetic field.

The Hamiltonian describing the quantum dot coupled
weakly to the left contact and strongly to the right contact (Fig.1)
has the form $H$$=$$H_0$$+$$H_L$$+$$H_R$$+H_C$, where
\begin{eqnarray}\nonumber
H_0=\sum_{k,\alpha}\epsilon_{k,\alpha}
c^\dagger_{k,\alpha}c_{k,\alpha} +\sum_{p,\alpha}\epsilon_{p,\alpha}
d^\dagger_{p,\alpha}d_{p,\alpha}\\
+\sum_{\alpha}\frac{v_{F,\alpha}}{2\pi}\!\!\!\int_{-\infty}^{\infty}\!\!\!\!\!\!\left\{\pi^2[\Pi_\alpha
(x,t)]^2+[\partial_x\phi_\alpha (x,t)]^2\right\}dx \label{h0}
\end{eqnarray}
describes a non-interacting part, $c$ denotes the electrons in the
left lead, $d$ stands for the electrons in the dot. Here
$\alpha=\uparrow,\downarrow$, $\phi_\alpha$ is a bosonization
displacement operator describing a transport through the QPC with
$\Pi_\alpha$ is conjugated momentum $[\phi_\alpha,
\Pi_{\alpha'}]=i\delta(x-x')\delta_{\alpha\alpha'}$. The Hamiltonian
$H_L$ describes the tunneling from the left (hot) lead to the dot
($t_{k,p,\alpha}$ is a tunnel amplitude)
\begin{eqnarray}
H_{L}=\sum_{k,p,\alpha}(t_{k,p,\alpha}
c^\dagger_{k,\alpha}d_{p,\alpha}F+\text{h.c.}). \label{hl}
\end{eqnarray}
The Hamiltonian $H_R$ accounts for the backward scattering in the
QPC, $r_\alpha$, are reflection amplitudes for
$\uparrow$$,$$\downarrow$
\begin{eqnarray}
H_{R}=-\frac{D}{\pi}\sum_{\alpha}|r_\alpha|\cos[2\phi_\alpha(0,t)].
\label{hr}\end{eqnarray}
\begin{figure}[t]
\includegraphics[width=5.0cm,angle=90]{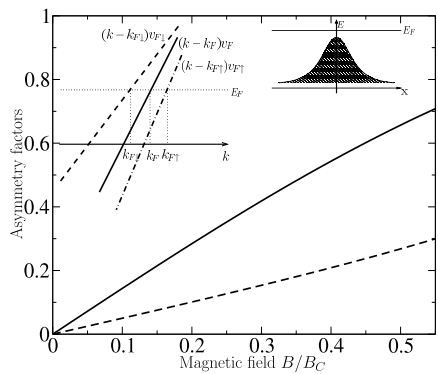}
\caption{Main frame: asymmetry factor (AF) for the reflection
amplitudes
$P_r=\left|\frac{|r_\uparrow|-|r_\downarrow|}{|r_\uparrow|+|r_\downarrow|}\right|$
of the QPC as a function of the magnetic field (solid line), AF for
the Fermi velocities
$P_v=\left|\frac{v_{F\uparrow}-v_{F\downarrow}}{v_{F\uparrow}+v_{F\downarrow}}\right|$
(dashed line). Insert 1: the Zeeman splitting of 1-d fermions
spectrum near the right Fermi-point. Insert 2: the coordinate
dependence of the energy in the vicinity of the QPC. The barrier is
depicted by shaded area.} \label{f.2}
\end{figure}
The modified form of the tunnel Hamiltonian takes into account the
change of the electron number $\hat n$ in the dot coming from the
left lead. For this sake the charge-lowering operator $\hat F$
satisfying the commutation relations $[\hat F,\hat n]=\hat F$ is
introduced in (\ref{hl}) (see details in \cite{PhysRevB.51.1743},
\cite{Averin90}). We treat the problem in the lowest order of tunnel
Hamiltonian and in linear response with respect to $\Delta T$.

The Hamiltonian $H_C$ describes the Coulomb interaction in the dot,
placed at coordinate $x$$=$$0$
\begin{eqnarray}
H_C=E_C\left[\sum_\alpha
c_\alpha^\dagger(0)c_\alpha(0)+\frac{1}{\pi}\sum_\alpha\phi_\alpha(0,t)-N(V_g)\right]^2.
\label{charging}
\end{eqnarray}
The electric- and thermo-currents through the dot are expressed in
this approximation in terms of Matsubara's Green Function (GF)
\begin{eqnarray}
\mathcal G(\tau)=-\sum_{p,p'\alpha}\langle T_\tau \hat
d_{p\alpha}(\tau)\hat F(\tau)\hat F^\dagger(0)\hat
d^\dagger_{p'\alpha}\rangle~.
\end{eqnarray}
Since the operators $\hat d$ and $\hat F$ are decoupled, the GF is
factorized into $\mathcal
G$$($$\tau$$)$$=$$G_0$$($$\tau$$)$$K($$\tau$$)$, with
$G_0(\tau)$$=$$-$$\nu_0$$\pi$$T$$/$$\sin$$($$\pi$$T$$\tau$$)$ being
the free electrons GF, $\nu_0$ is the density of states in the dot
without interaction and $K(\tau)$$=$$\langle$$T_\tau$$ \hat
F(\tau)$$ \hat F^\dagger$$(0)$$\rangle$ accounts for interaction
effects. The electric conductance \cite{FM_theory} is given by
\begin{eqnarray} G=\frac{G_L\pi T}{2}
\int_{-\infty}^{\infty}\frac{K\left(\frac{1}{2T}+it\right)}{\cosh^2(\pi
Tt)}dt ~. \label{ther_cond3}\end{eqnarray} The thermo-conductance
casts the form \cite{MA_theory}
\begin{eqnarray}
\!\!\!\!G_T=-\frac{i\pi^2}{2}\frac{G_L
T}{e}\int_{-\infty}^{\infty}\frac{\sinh(\pi T t)}{\cosh^3(\pi
Tt)}K\left(\frac{1}{2T}+it\right)dt ~. \label{cond3}\end{eqnarray}
Here $G_L$$\ll$$e^2$$/$$h$ denotes the tunnel conductance of the
left barrier calculated ignoring influence of the dot.

In order to calculate the thermoelectric coefficients for the model
(\ref{h0}-\ref{charging}) we generalize AM theory for the case of
finite magnetic field. The detailed calculations will be published
elsewhere \cite{nkk_future}. New effects reported in present Letter
and missing in AM theory appear due to asymmetry of the point
contact reflection amplitudes (Fig.2, Insert 2) in a presence of
Zeeman splitting (cf. \cite{PhysRevB.64.161302}). This asymmetry, in
turn, leads to the asymmetry of the channels in 2CK. Besides, the
magnetic field lifts out the spin-charge separation characteristic
for the unpolarized models \cite{Giamarchi_book}. Since the system
possesses the particle-hole symmetry at a perfect transmission, the
leading effects determining behavior of the thermoelectric
coefficients appear only at non-zero backward scattering. This is
why the major effect is related to asymmetry of $|r|$'s. Thus, we
are allowed to disregard the effects of the Fermi-velocities
difference in a full agreement with standard bosonization scheme
\cite{Giamarchi_book}. The effects of spectrum curvature responsible
for particle-hole asymmetry are crucial for understanding of Coulomb
drag effect in the Luttinger Liquids \cite{Glazman}. The finite
curvature effects, however result in sub-leading corrections to TP
for the AM model while leading effect comes from the scattering
potential asymmetry. Moreover, the saturation of
$|r_\downarrow|$$\to$$1$ occurs at magnetic fields
$B^\ast$$\ll$$B_C$ (see left insert on Fig.3), $B_C$ is defined as a
field corresponding to full polarization of the QPC. The channel
$\downarrow$ becomes completely reflecting and therefore does not
contribute to the transport. It explains the crossover from
$\sim$$|r|^2$ for small $B$ (\ref{sf1}) to $\sim$$|r|$ for
$B$$>$$B^\ast$ (\ref{sl1}) in TP.

In a spirit of AM theory \cite{MA_theory} we first calculate the
leading non-trivial correction to the thermo-conductance $G_T$ and
the thermopower $S$ for the case of slightly asymmetric reflection
amplitudes. The equation (\ref{sf1}) is modified by a replacement
$|r|^2\to |r_\uparrow r_\downarrow|$. Thus, all perturbative
corrections depend on a symmetric
$s=|r_\uparrow|+|r_\downarrow|\approx 2|r_0|$ and anti-symmetric
$a=||r_\downarrow|-|r_\uparrow||\sim |r_0 B/B_C|$ combinations of
the reflection amplitudes. Here $|r_0|$ stands for the reflection
amplitude at $B$$=$$0$. Proceeding with the higher-order
perturbative corrections to thermo-conductance and TP we notice
existence of a new energy scale (similar to $\Gamma$ in (\ref{sf2}))
depending on external magnetic field \cite{nkk_future}.
\begin{figure}[t]
\includegraphics[width=6.0cm,angle=-90]{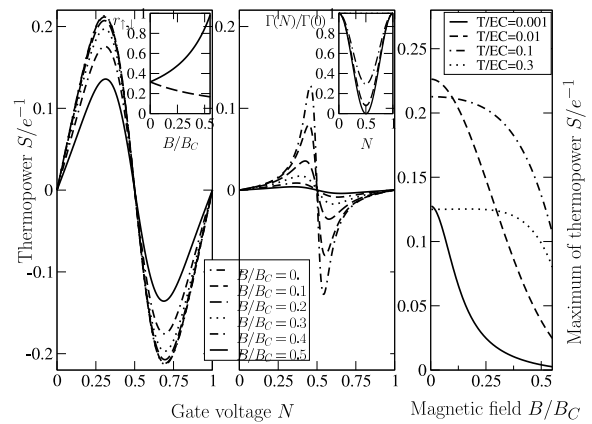}
\caption{Left and central panels: TP as a function of the gate
voltage at different magnetic fields. The curves are obtained
numerically for $T$$/$$E_C$$=$$0.1$ (left panel) and
$T/E_C$$=$$0.001$ (central panel). For both cases $|r_0|^2=0.1$. The
insert on the left panel shows the magnetic field dependence for the
down (solid line) and up (dashed line) reflection amplitudes. The
insert on the central panel shows the gate voltage dependence of
$\Gamma$. Right panel: maximum of TP in a Coulomb window at various
fixed temperatures as the function of magnetic field.}\label{f.3}
\end{figure}

More efficient way to prove emergence of new energy scale is to map
the model (\ref{h0}-\ref{charging}) onto effective Anderson model
\cite{PhysRevB.51.1743}. This mapping, being non-perturbative in $s$
and $a$ accounts for nontrivial low-frequency dynamics of the spin
modes. The channel asymmetry $a$ leads to nontrivial contribution to
the Kondo resonance $\Gamma$ in the vicinity of Coulomb peaks
\begin{eqnarray}
\Gamma= \frac{2\gamma E_C}{\pi^2}\left[s^2\cos^2\left(\pi N\right)+
a^2\sin^2\left(\pi N\right)\right].\label{gg}
\end{eqnarray}
As it is shown in the central insert of Fig. 3, $\Gamma$ is gapped
at the Coulomb peaks with the gap
$T_{min}$$=$$\Gamma_{min}$$\sim$$a^2$$E_C$ depending on the magnetic
field through $a$. As a result, the thermal and electric
conductances acquire a form
\begin{eqnarray}\nonumber
&&\!\!\!\!\!\!\!G_T=-\frac{1}{12\pi}\frac{G_L}{e}\left(s^2-a^2\right)\frac{T}{E_C}\ln\frac{E_C}{T+\Gamma}\sin(2\pi
N) F_1\left(\frac{\Gamma}{T}\right)\\
&&\!\!\!\!\!\!\!G=\frac{G_L T}{4\gamma
E_C}F_2\left(\frac{\Gamma}{T}\right). \label{ge}
\end{eqnarray}
The functions $F_1(x)$ and $F_2(x)$ are defined in \cite{com1},
$\gamma$$\approx$$1.78$. The equations ({\ref{ge}) allow regular
expansion at $T$$\ll$$T_{min}$. In the Coulomb valleys the AM
results are reproduced. The effects of magnetic field in the valley
regime manifest themselves in reducing the TP and restoration of the
sinusoidal shape of the TP curve. In the vicinity of Coulomb peaks
and at sufficiently low temperatures $T$$\ll$$T_{min}$ the equations
(\ref{ge}) cast the form \vspace*{-3mm}
\begin{eqnarray}\nonumber\label{key}
&&G_T=-\frac{2\pi^3}{15}\frac{G_L}{e}\left(s^2-a^2\right)\frac{T^3}{E_C \Gamma_a^2}\ln\frac{E_C}{\Gamma_a}\sin(2\pi N)~,\nonumber\\
&&G=\frac{\pi^2 G_L}{6\gamma}\frac{T^2}{E_C \Gamma_a}~,\\
&&S=
-\frac{4\pi\gamma}{5e}\frac{T}{\Gamma_a}(s^2-a^2)\ln\left[\frac{E_C}{\Gamma_a}\right]\sin\left(2\pi
N\right)\nonumber,
\end{eqnarray}
where $\Gamma_a=\frac{2\gamma E_C}{\pi^2}a^2\sin^2(\pi N)$. Thus,
the re-scaled FL regime is restored at low temperatures
$T$$\ll$$\Gamma_a$$($$N$$\to$$1/2$$)$.

The equations (\ref{gg}-\ref{key}) represent the central result of
this Letter. The new energy scale $T_{min}$ strongly depends on
magnetic field and manifests itself by opening a gap at the Coulomb
peaks. Appearance of this scale \cite{PhysRevB.64.161302} leads to
the restoration of the FL thermo- and electric- transport
properties. However, proximity to NFL regime at $B$$\ll$$B_C$
results in strong enhancement of the TP. When the channel
$\downarrow$ becomes fully reflecting, the equation (\ref{key})
reduces to 1CK result (\ref{sl1}). The curves on right panel of
Fig.3 are obtained as numerical solution of (\ref{ge}) and show the
evolution of TP amplitude in the interval of gate voltages $N$,
$N$$+$$1$ at fixed temperatures as a function of magnetic field. The
behavior of TP at temperatures $T>$$T_{max}$$\sim$$s^2$$E_C$ is
trivial and described by Eq.\ref{sf1}. The crossover from NFL to FL
is seen at the intermediate range of temperatures
$T_{min}$$<$$T$$<$$T_{max}$. Thus, the magnetic field stabilizes the
FL thermoelectric properties.

We conclude that the external magnetic field applied to the SET in
the strong coupling regime is responsible for the NFL-to-FL change
of transport properties and provides efficient mechanism for the
restoration of the FL behavior in the thermoelectric transport. The
applied to SET parallel magnetic field $B$ results in the channel
up/down asymmetry and thus changes the universality class from
two-channel Kondo to single-channel Kondo regimes. Another
possibility of the 2CK suppression is associated with a finite
source-drain voltage or noise, which is equivalent to the Zeeman
effect for the conventional Kondo systems. We also expect
restoration of the FL transport for the out-of-equilibrium SETs.

We are grateful to K.A. Matveev for illuminating discussions and
L.W.Molenkamp for drawing our attention to thermoelectric transport
through nanostructures. We appreciate valuable comments by
Y.Alhassid, B.L.Altshuler, K.Behnia, K.Kikoin and N.Prokof'ev. We
acknowledge support through INT-09-2b program and warm hospitality
of the University of Washington, Seattle where part of this work was
performed. TKTN visit to Seattle was supported through ICAM-I2CAM
grant.


\vspace*{-6mm}
\begin{thebibliography}{99}
\vspace*{-6mm}
\bibitem{Staring93} A. A. M. Staring et al,
Europhys. Lett, {\bf 22}, 57 (1993).

\bibitem{Dzurak97} A.S.Dzurak et al,
Phys. Rev. B {\bf 55}, R10197 (1997).

\bibitem{PhysRevLett.95.176602}
R. Scheibner et al,
Phys. Rev. Lett. {\bf 95},  176602  (2005).

\bibitem{PhysRevB.75.041301(R)}
R.~Scheibner et al,
Phys. Rev. B {\bf 75}, 041301  (2007).

\bibitem{PhysRevB.46.9667}
C.~W.~J. Beenakker and A.~A.~M. Staring, Phys. Rev. B {\bf 46},
9667  (1992).

\bibitem{MA_theory} A.~V. Andreev and K.~A. Matveev, Phys. Rev. Lett. {\bf 86},  280
(2001), K.~A. Matveev and A.~V. Andreev, Phys. Rev. B {\bf 66},
045301 (2002).

\bibitem{PhysRevB.65.115332}
M.Turek, K.A. Matveev, Phys.Rev.B {\bf 65}, 115332 (2002).

\bibitem{EuroPhysLett.56.576}
D. Boese and R. Fazio, Europhys. Lett. {\bf 56},  576  (2001).

\bibitem{Goldhaber} R. M. Potok et al,
Nature (London) {\bf 446}, 167 (2007).


\bibitem{Cutler69} M.~Cutler and N.F.~Mott, Phys. Rev {\bf 181}, 1336
(1969).

\bibitem{Moller98} S. M\"oller et al,
Phys. Rev. Lett. {\bf 81}, 5197 (1998).

\bibitem{PhysRevB.51.1743}
K.~A. Matveev, Phys. Rev. B {\bf 51},  1743  (1995).

\bibitem{FM_theory}
A. Furusaki and K.~A. Matveev, Phys. Rev. Lett. {\bf 75},  709
(1995);  Phys. Rev. B {\bf 52}, 16 676 (1995).

\bibitem{Flensberg} K. Flensberg, Phys. Rev. B 48, 11 156 (1993).


\bibitem{Houten92} H.van Houten et al,
Semicond.Sci.Technol. {\bf 7},B215 (1992)

\bibitem{cm1} The CB in almost open QD with finite mean level
spacing has been considered in I. L. Aleiner and L. I. Glazman,
Phys. Rev. B {\bf 57}, 9608 (1998).

\bibitem{com00} The magnetic field $B$ is assumed
parallel to the plane of 2 dimensional electron gas. No orbital
effects are discussed.


\bibitem{com} Similar behaviour is known in heavy-fermion strongly
correlated electron systems. The energy scale $T^\ast\ll \epsilon_F$
is attributed to "heavy" electron's mass.

\bibitem{Ong03} Y. Wang et at,
Nature (London) {\bf 423}, 425 (2003)



\bibitem{com1} The function $f(x)=F_1(x)/F_2(x)$ is defined by two
functions $F_1(x)=\int_{0}^{\infty} dz
z^2(z^2+\pi^2)/[(z^2+x^2)\cosh^2(z/2)]$ and
$F_2(x)=x\int_{0}^{\infty} dz (z^2+\pi^2)/[(z^2+x^2)\cosh^2(z/2)]$.

\bibitem{NozBla80} P.Nozieres and A.Blandin, J.Phys (Paris) {\bf 41}, 193
(1980)

\bibitem{Affleck91} A.W.W. Ludwig and I. Affleck,   Phys. Rev. Lett. {\bf 67}, 3160 (1991).


\bibitem{com2} It has been shown in \cite{PhysRevB.51.1743} that the strong coupling
fixed point of the model (\ref{h0}-\ref{charging}) is identical to
the strong coupling fixed point of the two-chanel Kondo model. Note,
that the solution \cite{PhysRevB.51.1743} is valid only on the
separatrix $|r_\uparrow|$$=$$|r_\downarrow|$, being unstable
everywhere else.


\bibitem{PhysRevB.64.161302}
K. Le~Hur, Phys. Rev. B {\bf 64},  161302(R)  (2001), K. Le~Hur and
G. Seelig, Phys. Rev. B {\bf 65}, 165338 (2002).


\bibitem{Averin90} D.V. Averin and Yu.V. Nazarov,  Phys. Rev. Lett. {\bf 65}, 2446
(1990).


\bibitem{nkk_future} T. K. T. Nguyen et al, (to be published)


\bibitem{Giamarchi_book} T. Giamarchi, {\it Quantum physics in one dimension} (Oxford University Press, 2004).

\bibitem{Glazman} M. Pustilnik et al,
Phys. Rev. Lett. {\bf 91}, 126805 (2003).


\end{thebibliography}
\end{document}